# Nonreciprocal subdiffraction imaging with staggered gyromagnetic photonic crystals

Rui Ding[1], Tianshu Zhang[2], Jiarui Yu[1], Namitha Nandakumar[2], Quanlong Yang[3], Mudi Wang[1,*], D.Y. Wang[2,*]

[1]Key Laboratory of Artificial Micro- and Nanostructures of Ministry of Education and School of Physics and Technology, Wuhan University, Wuhan, China

[2]Optoelectronics Research Centre, University of Southampton, Southampton SO17 1BJ, UK

[3]School of Physics, Central South University, Changsha 410083, China

*Corresponding authors: mudiwang@whu.edu.cn; d.y.wang@soton.ac.uk

**Abstract**

Subdiffraction imaging serves as a novel technique to enhance the optical imaging resolution, where a backscattering-free approach remains so far unavailable. Here, we theoretically predict and experimentally demonstrate that the gyromagnetic photonic crystals applied with staggered magnetic fields support nonreciprocal light propagation beyond the diffraction limit. Broadband subdiffraction imaging was experimentally observed to span the frequency window traversing a pair of spectrally separated Dirac points, where near-flat equi-frequency contours (EFCs) emerged while experiencing an enforced shape transition. Our findings establish a practical paradigm for constructing backscattering-immune super-resolution imaging systems based on topological photonic crystal platforms.

## Introduction

Topological photonics [1-3] provides a robust paradigm for manipulating light propagation beyond conventional optical limits. In particular, various topological photonic phases have been discovered to support reflection-free transport, such as the chiral edge states in two-dimensional Chern insulators [4-7], Fermi-arcs in three-dimensional topological Weyl semimetals [8-16], and recently developed antichiral transport in staggered gyromagnetic photonic crystals (SGPCs) [17,18]. Such robust light propagation empowers high-performance photonic devices with suppressed energy loss.

Subdiffraction imaging constitutes a unique physical effect that overcomes the wavelength-imposed bound on optical imaging resolution, and it has garnered substantial research interest [19-26]. Breaking this diffraction barrier requires capturing large wavevector light components outside the light cone, which encodes spatial features smaller than the operating optical wavelength. The full set of propagating wavevectors sustained by an optical medium is geometrically characterized by equi-frequency contours (EFCs) defined within momentum space. For ordinary homogeneous dielectrics, EFCs take the form of closed circular or elliptical loops in two-dimensional momentum space. By contrast, materials featuring permittivity or permeability tensors with mixed positive and negative diagonal entries support open hyperbolic EFCs [24], exemplified by layered metamaterials with alternating metallic and dielectric slabs. These so-called hyperbolic media support infinite large wavevector components of light, thereby enabling key functional devices such as superlenses [27].

Recent studies have explored the utilization of topological photonic modes for subwavelength imaging [28], demonstrating that flat surface-state EFCs in topological metamaterials can facilitate imaging beyond the diffraction limit. Nevertheless, unidirectional and nonreciprocal imaging remains unrealized in such systems due to inherent fundamental symmetry constraints. Here, we propose a strategy to achieve nonreciprocal subdiffraction imaging by exploiting the bulk propagation modes of SGPCs, wherein both time-reversal symmetry and inversion symmetry are effectively broken. We reveal that the two Dirac points at K/K' valleys induce morphological transitions of EFCs in momentum space, with near-flat EFCs emerging as an intermediate state that enables robust subdiffraction imaging. Furthermore, the symmetry breaking and frequency misalignment of Dirac points eliminate backscattering pathways entirely, endowing the proposed imaging scheme with full nonreciprocity. By fabricating the waveguide-like two-dimensional SGPCs sample, such nonreciprocal subdiffraction imaging was experimentally validated.

**Staggered gyromagnetic photonic crystals**

The SGPCs platform is constructed with a rhombohedral lattice consisting of two gyromagnetic sublattices (A and B), composed of yttrium iron garnet (YIG) rods embedded in an air background in Fig. 1(a). The two sublattices are subjected to spatially staggered opposite out-of-plane magnetic fields $B_{\pm}$, breaking both time-reversal and inversion symmetry. The structural parameters are fixed as lattice constant $a$ = 10 mm and rod radius $r = 0.2a$. The YIG material presents a constant permittivity of $\varepsilon_r$ = 13.8, but it carries a dispersive permeability tensor under magnetic field as [29]:

$$\mu_{\pm} = \begin{bmatrix} \mu_r & \pm i\mu_c & 0 \\ \mp i\mu_c & \mu_r & 0 \\ 0 & 0 & 1 \end{bmatrix},$$

where $\mu_r = 1 + \frac{(\omega_0 + i\alpha\omega)\omega_m}{(\omega_0 + i\alpha\omega)^2 - \omega^2}$, $\mu_c = \frac{\omega\omega_m}{(\omega_0 + i\alpha\omega)^2 - \omega^2}$, $\omega_0 = \mu_0\gamma H_0$, and $\omega_m = \mu_0\gamma M_s$. The material parameters are [30]: gyromagnetic ratio $\gamma = 2.8\times10^{10}$ $T^{-1}s^{-1}$, saturation magnetization $\mu_0 M_s$ = 1850 G, and damping factor $\alpha \approx 0.005$.

We consider an effective magnetic field of $|B_{\pm}| = \mu_0 H_0 \approx$ 850 G, resulting into a magnetic resonance around $f_0 \approx$ 2.3 GHz, being away from the frequency range of interest. The material dispersion can thus be neglected, and constant permeability tensor elements are taken as $\mu_r$ = 0.79 and $\mu_c$ = 0.71 for the numerical calculations. Adopting these parameters, the band structures are calculated in Fig. 1(b) for the transverse magnetic modes in the SGPCs by using the eigenmode solver of COMSOL Multiphysics. The first Brillouin Zone (BZ) information is presented in Fig. 1(a) with the high symmetry points marked in red. In the band structures, since both the time-reversal symmetry and the inversion symmetry were broken for the SGPCs, the two Dirac points at K and K' positions occur at distinct frequencies of $f_K \approx$ 9 GHz and $f_{K'} \approx$ 8 GHz, respectively.

As the SGPCs faithfully reproduce the spectrally split Dirac points in the modified Haldane model [4, 17], the model Hamiltonian $H = t_1 \sum_{\langle i,j \rangle} c_i^\dagger c_j + t_2 \sum_{\langle\langle i,j \rangle\rangle} e^{-i\phi} c_i^\dagger c_j$ can be adopted here to predict the EFCs' evolution with frequency. The three-dimensional band dispersions are hence calculated for a set of parameters $t_1$ = 1, $t_2$ = 0.15, and $\phi = \pi/2$, as shown in the Fig. 1(c), where the EFCs are successively enclosing the Γ-K'-K-Γ points as frequency increases. Such change of loop centre forces the EFCs to become open arcs at the intermedia stats, where near-flat EFCs emerges, as can be recognized in Fig. 1(c). These EFCs are the key features to explore in the practical SGPCs material in later section.

To build the SGPCs, detailed sample design is illustrated schematically in Fig. 1(d), where yttrium iron garnet (YIG) rods are sandwiched between two copper plates. Permanent magnets are installed on the top and bottom sides of the YIG rods and embedded within the metallic plates. Owing to such top and bottom metallic claddings, the constructed SGPCs structure support only transverse magnetic modes, which is consistent with the mode configuration adopted for the band structure calculations in Fig. 1(b). Alternating magnetic field orientations can be realized by arranging the embedded magnets in a staggered configuration. To facilitate direct internal field measurement within the SGPCs region, periodic holes are drilled through the metallic plates to form a two-dimensional grid of probing points.

**Observation of spectrally split Dirac points and EFCs shape transition.**

To experimentally investigate the transport property of the SGPCs, we take a measurement configuration as shown in Fig. 2(a). The grey dots interlacing the YIG rods in red/blue indicate the drilled hole positions in the metallic plate, which serve as available probing positions. The step widths for measurements are $\Delta_x = a/2$ and $\Delta_y = a/2\sqrt{3}$ along x and y directions, respectively.

Firstly, we measured the photonic band projections (along $k_x$) in the SGPCs by placing a point-source to the middle of the sample and probing the electromagnetic field distribution along the horizontal dashed line in Fig. 2(a). By performing Fourier transform on the measured spatial field profile, the wavevector-frequency dispersion of the propagating modes in the SGPCs can be retrieved. The calculation and experimental results on the projected bands were shown in Fig. 2(b) and (c), respectively, where the two Dirac points at the K and K' positions were experimentally verified at $f_K^{Exp} \approx 8.7$ GHz and $f_{K'}^{Exp} \approx 8$ GHz.

We now proceed to study the two-dimensional bulk band EFCs for the SGPCs. The existence of Dirac points at the K and K' position and their frequency misalignment lead to interesting shape transitions of EFCs in the first BZ, similar to the modal prediction in Fig. 1(c). To illustrate such transition process, we numerically calculate the EFCs for several frequencies as shown in Fig. 2(d). At the frequency of $f_{K'} = 8$ GHz, point singularities can be found at K' position corresponding to the Dirac points, and closed contours are surrounding the K positions. As frequency gradually increases towards $f_K = 9$ GHz, the EFCs around K' point smoothly expand, while the EFCs at K positions gradually shrink into point singularities. The group velocity directions can be examined for the EFCs at $f = 9$ GHz, as being marked by the red arrows. These group velocity directions indicate that the light propagation along positive x-direction (and the other two equivalent directions due to crystalline rotation symmetry) is supported. Importantly, because of the frequency misalignment between Dirac points at K and K’, there is no back-reflection channel for the EFCs in the BZ, so these propagations are fully nonreciprocal. As frequency further increase, the closed contours encircling the K positions will further expand and evolve into near-flat arcs as shown in the last panel of Fig. 2(d) for $f = 9.3$ GHz. Such EFCs subsequently turn into contours enclosing the Γ point due to the band folding at K positions. Since the width of these EFCs along $k_y$-direction surpasses the light cone (indicated by the brown circle in the BZ centre), they will support light propagation beyond the diffraction limit.

To experimentally validate the EFCs evolution, we placed a point source at the SGPCs center (marked by the red star in Fig. 2(a)) and characterized the electromagnetic field distribution in the rectangular region shown in Fig. 2(a). Fourier transformation of the measured near-field pattern yields the momentum-space EFCs in Fig. 2(e), which verifies the morphological transition of EFCs. Notably, the slight frequency shifts of the experimentally measured Dirac points and EFCs are primarily attributed to material dispersion and discrepancies in effective magnetic field strength between experimental measurements and numerical simulations.

**Experimental demonstration of nonreciprocal subdiffraction imaging**

The near-flat EFCs observed in Fig. 2(d-e) serve to provide uniform group velocities for a continuum collection of wavevectors beyond the regime of light cone, which laid out the foundation for subdiffraction imaging. We firstly consider the configuration of a single point-source in the SGPCs with no material loss, and perform numerical simulations to visualize light propagation, as depicted in Fig. 3(a). It can be easily seen that light beams were excited to propagate towards three equivalent K-directions.

To experimentally verify such light propagation, we took a line-section as indicated in Fig. 3(a) and measured the electromagnetic field distribution along it. The measured electric field distributions are presented in Fig. 3(b) for a series of frequencies, where a beam width of $w \approx 3\Delta_y = 8.66$ mm has been observed. This corresponds to $w \approx 0.26\lambda$ at $f = 9$ GHz, presenting a subwavelength beam profile.

As the near-flat EFCs emerge in the process of smooth shape transitions, the collimated subwavelength beam can be expected from a range of nearby frequencies. This is verified by the well-maintained beam profile in Fig. 3(b) for the frequencies from $f = 8.6$ GHz to $f = 10$ GHz. The spectra manifestation of beam profile is more clearly presented in Fig. 3(c), where the position of the point-source is marked by a star, and the subwavelength beam width was observed to extend from $f \approx 8.5$ GHz to $f \approx 11$ GHz.

To demonstrate the elementary idea of subdiffraction imaging, we set up two point-sources with subwavelength distances of $d = 4\Delta y \approx 11.55$ mm and examine whether the two sources can be distinguished while the propagation of light in the SGPCs. Full-wave simulations were firstly performed to reveal the electric field $|E_z|$ distribution for $f = 8.9$ GHz in Fig. 3(d). As can be seen, two beams were excited to propagate along x-direction, and their beam profiles were well maintained while propagating in the SGPCs to distinguish the two point-sources. It should be noted that due to the lack of three-fold rotation symmetry in the point-source configuration, not all equivalent K-directions were excited.

To experimentally verify the subdiffraction imaging on two point-sources with the SGPCs, we measured the electric field distribution inside an area marked by the box in Fig. 3(d). The experimental results are shown in Fig. 3(e), where two beams can be found to propagate along x-direction, as indicated by the white boxes.

Since only those discrete grid points avoiding the YIG rods positions are available for experimentally probing the field, the measured electric field patten in Fig. 3(e) shows dependence on the even or odd column. We thus take two line-sections representing each case in Fig. 3(e) to examine the beam profile and spectra performance of the subdiffraction imaging. For the blue dashed line in Fig. 3(e), the electric field $|E_z|$ distributions and spectra variance are shown in Fig. 3(f) and (g), respectively. Two peaks separated by $\delta = 4\Delta_y$ can be found in Fig. 3(f) that clearly distinguish the two beams excited by the point-sources, and a clear image of the two point-sources can be found in Fig. 3(g) within the spectra range from $f \approx 8$ GHz to $f \approx 11$ GHz. Similarly, the results for the pink dashed line in Fig. 3(e) are shown in Fig. 3(h) and (i), where the two peaks are now with a nearer distance of $\delta = 2\Delta_y$, but still observable within the spectra range from $f \approx 8.5$ GHz to $f \approx 11$ GHz. These results demonstrate that the two point-sources separated by $d = 4\Delta_y \approx 11.55$ mm can be experimentally distinguished in the SGPCs, which marks an imaging capability of $\eta = d/\lambda \approx 0.35$ referencing to $f = 9$ GHz.

**Conclusion**

In conclusion, this work reveals that SGPCs exhibit near-flat EFCs, which support nonreciprocal light propagation and enable effective diffraction-limit breaking. Experimental results validate the broadband nonreciprocal subdiffraction imaging capability of the proposed system, achieving a high imaging resolution of $\eta = d/\lambda \approx 0.35$ at $f = 9$ GHz. This study uncovers a new physical mechanism for nonreciprocal super-resolution imaging and offers a feasible platform for the development of topological nonreciprocal imaging devices, high-resolution optical systems, and robust photonic integrated circuits.

**Acknowledgement**
This work is supported by the Royal Society research grant: RGS\R2\252056. D.Y. W. acknowledges support from the Anniversary Fellowship of the University of Southampton. M. W. acknowledges the National Natural Science Foundation of China Grant No. 12504251.

**References**

[1] M.Z. Hasan, C.L. Kane, Colloquium: Topological insulators, Reviews of Modern Physics, 82 (2010) 3045-3067.

[2] X.-L. Qi, S.-C. Zhang, Topological insulators and superconductors, Reviews of Modern Physics, 83 (2011) 1057-1110.

[3] L. Lu, J.D. Joannopoulos, M. Soljačić, Topological photonics, Nature Photonics, 8 (2014) 821-829.

[4] F.D.M. Haldane, Model for a Quantum Hall Effect without Landau Levels: Condensed-Matter Realization of the "Parity Anomaly", Physical Review Letters, 61 (1988) 2015-2018.

[5] F.D.M. Haldane, S. Raghu, Possible Realization of Directional Optical Waveguides in Photonic Crystals with Broken Time-Reversal Symmetry, Physical Review Letters, 100 (2008) 013904.

[6] Z. Wang, Y.D. Chong, J.D. Joannopoulos, M. Soljačić, Reflection-Free One-Way Edge Modes in a Gyromagnetic Photonic Crystal, Physical Review Letters, 100 (2008) 013905.

[7] Z. Wang, Y. Chong, J.D. Joannopoulos, M. Soljačić, Observation of unidirectional backscattering-immune topological electromagnetic states, Nature, 461 (2009) 772-775.

[8] X. Wan, A.M. Turner, A. Vishwanath, S.Y. Savrasov, Topological semimetal and Fermi-arc surface states in the electronic structure of pyrochlore iridates, Physical Review B, 83 (2011) 205101.

[9] L. Lu, L. Fu, J.D. Joannopoulos, M. Soljačić, Weyl points and line nodes in gyroid photonic crystals, Nature Photonics, 7 (2013) 294-299.

[10] L. Lu, Z. Wang, D. Ye, L. Ran, L. Fu, J.D. Joannopoulos, M. Soljačić, Experimental observation of Weyl points, Science, 349 (2015) 622-624.

[11] M. Xiao, W.-J. Chen, W.-Y. He, C.T. Chan, Synthetic gauge flux and Weyl points in acoustic systems, Nature Physics, 11 (2015) 920-924.

[12] W. Gao, B. Yang, M. Lawrence, F. Fang, B. Béri, S. Zhang, Photonic Weyl degeneracies in magnetized plasma, Nature Communications, 7 (2016) 12435.

[13] B. Yang, Q. Guo, B. Tremain, L.E. Barr, W. Gao, H. Liu, B. Béri, Y. Xiang, D. Fan, A.P. Hibbins, S. Zhang, Direct observation of topological surface-state arcs in photonic metamaterials, Nature Communications, 8 (2017) 97.

[14] B. Yang, Q. Guo, B. Tremain, R. Liu, L.E. Barr, Q. Yan, W. Gao, H. Liu, Y. Xiang, J. Chen, C. Fang, A. Hibbins, L. Lu, S. Zhang, Ideal Weyl points and helicoid surface states in artificial photonic crystal structures, Science, 359 (2018) 1013-1016.
[15] D. Wang, B. Yang, W. Gao, H. Jia, Q. Yang, X. Chen, M. Wei, C. Liu, M. Navarro-Cía, J. Han, W. Zhang, S. Zhang, Photonic Weyl points due to broken time-reversal symmetry in magnetized semiconductor, Nature Physics, 15 (2019) 1150-1155.
[16] D. Wang, H. Jia, Q. Yang, J. Hu, Z.Q. Zhang, C.T. Chan, Intrinsic Triple Degeneracy Point Bounded by Nodal Surfaces in Chiral Photonic Crystal, Physical Review Letters, 130 (2023) 203802.
[17] E. Colomés, M. Franz, Antichiral Edge States in a Modified Haldane Nanoribbon, Physical Review Letters, 120 (2018) 086603.
[18] P. Zhou, G.-G. Liu, Y. Yang, Y.-H. Hu, S. Ma, H. Xue, Q. Wang, L. Deng, B. Zhang, Observation of Photonic Antichiral Edge States, Physical Review Letters, 125 (2020) 263603.
[19] S. Zhang, Y. Xiong, G. Bartal, X. Yin, X. Zhang, Magnetized Plasma for Reconfigurable Subdiffraction Imaging, Physical Review Letters, 106 (2011) 243901.
[20] J.B. Pendry, Negative Refraction Makes a Perfect Lens, Physical Review Letters, 85 (2000) 3966-3969.
[21] N. Fang, H. Lee, C. Sun, X. Zhang, Sub-Diffraction-Limited Optical Imaging with a Silver Superlens, Science, 308 (2005) 534-537.
[22] T. Taubner, D. Korobkin, Y. Urzhumov, G. Shvets, R. Hillenbrand, Near-Field Microscopy Through a SiC Superlens, Science, 313 (2006) 1595-1595.
[23] J.B. Pendry, D. Schurig, D.R. Smith, Controlling Electromagnetic Fields, Science, 312 (2006) 1780-1782.
[24] D.R. Smith, D. Schurig, Electromagnetic Wave Propagation in Media with Indefinite Permittivity and Permeability Tensors, Physical Review Letters, 90 (2003) 077405.
[25] Z. Jacob, L.V. Alekseyev, E. Narimanov, Optical Hyperlens: Far-field imaging beyond the diffraction limit, Opt. Express, 14 (2006) 8247-8256.
[26] Z. Liu, H. Lee, Y. Xiong, C. Sun, X. Zhang, Far-Field Optical Hyperlens Magnifying Sub-Diffraction-Limited Objects, Science, 315 (2007) 1686-1686.
[27] A. Poddubny, I. Iorsh, P. Belov, Y. Kivshar, Hyperbolic metamaterials, Nature Photonics, 7 (2013) 948-957.
[28] D. Wang, B. Yang, R.-Y. Zhang, W.-J. Chen, Z.Q. Zhang, S. Zhang, C.T. Chan, Straight Photonic Nodal Lines with Quadrupole Berry Curvature Distribution and Superimaging "Fermi Arcs", Physical Review Letters, 129 (2022) 043602.

[29] D.M. Pozar, Microwave engineering, Microwave engineering2004.
[30] M. Wang, R.-Y. Zhang, L. Zhang, D. Wang, Q. Guo, Z.-Q. Zhang, C.T. Chan, Topological One-Way Large-Area Waveguide States in Magnetic Photonic Crystals, Physical Review Letters, 126 (2021) 067401.

**Figures**

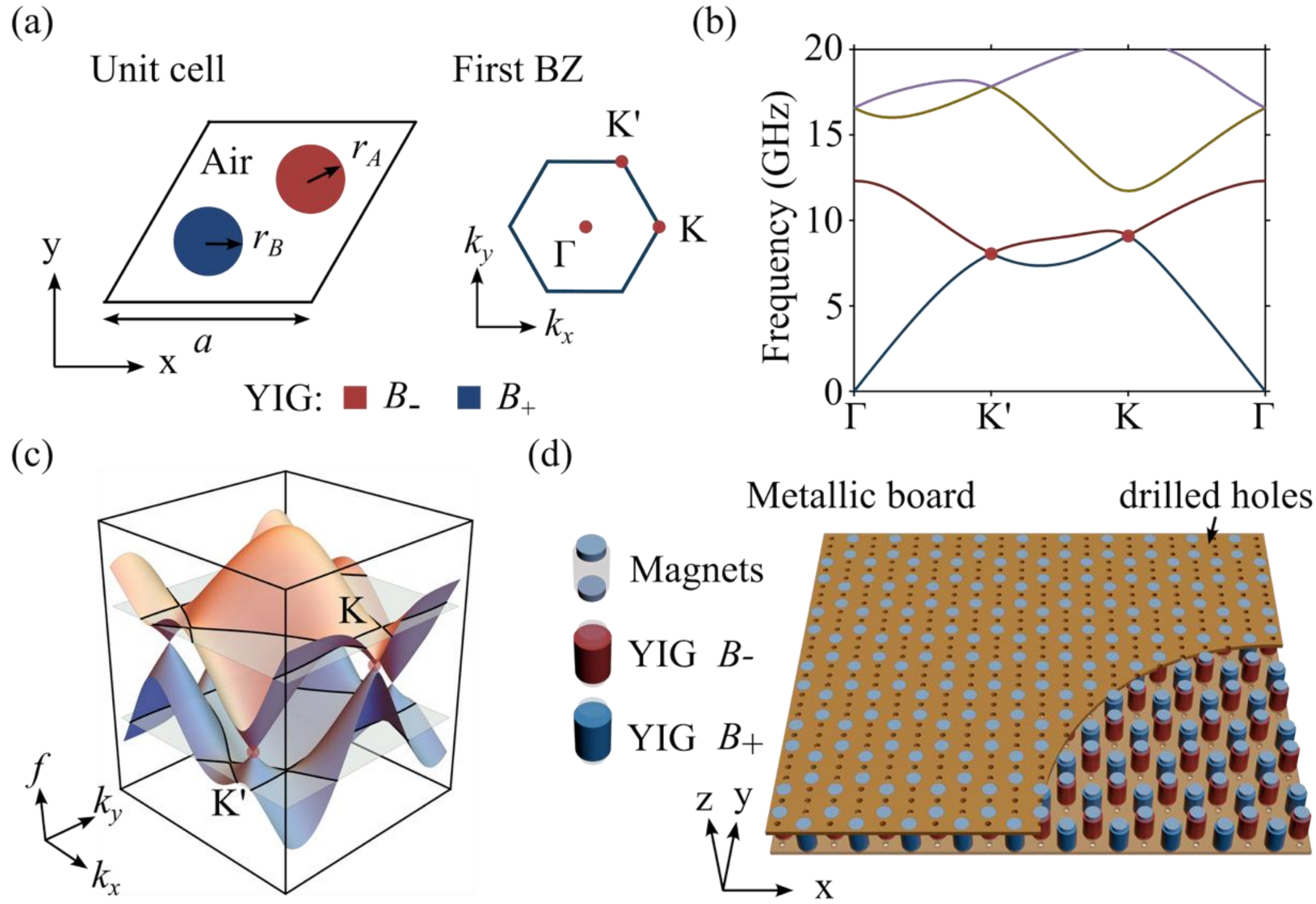

FIG. 1. Staggered gyromagnetic photonic crystals. (a) The unit cell and first Brillouin zone of the photonic crystal. (b) The photonic band structures with spectrally split Dirac points. (c) The three-dimensional band structures of modified Haldane model with near-flat EFCs being indicated. (d) The schematic of the sample design.

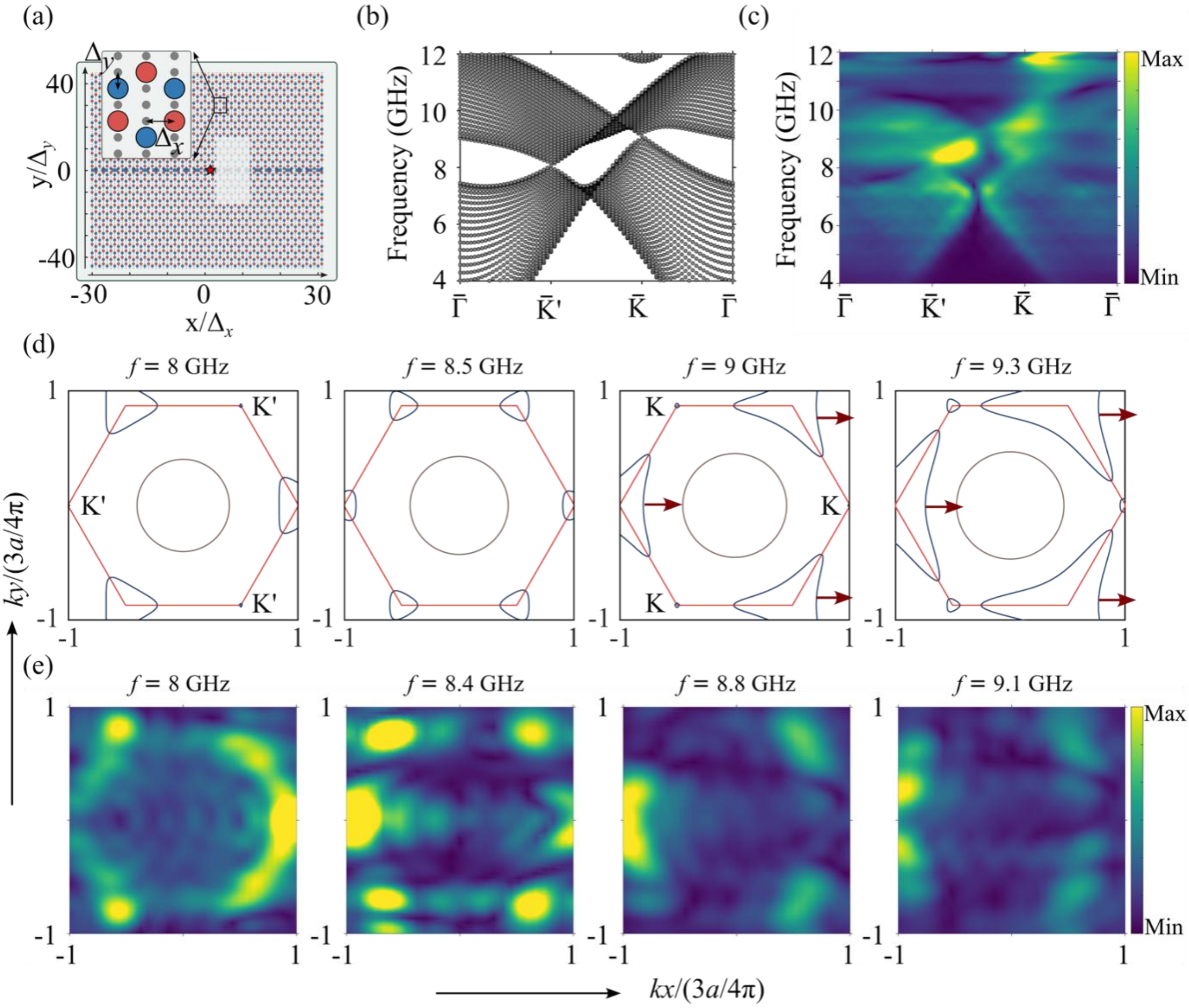

FIG. 2. Experimental measurements on the projected bands and EFCs. (a) Illustration of the experimental measurement configuration. (b) Calculated projected band dispersions along $k_x$-direction. (c) Experimentally measured projected bands. (d) Calculated EFCs for different frequencies. (e) Experimentally measured EFCs.

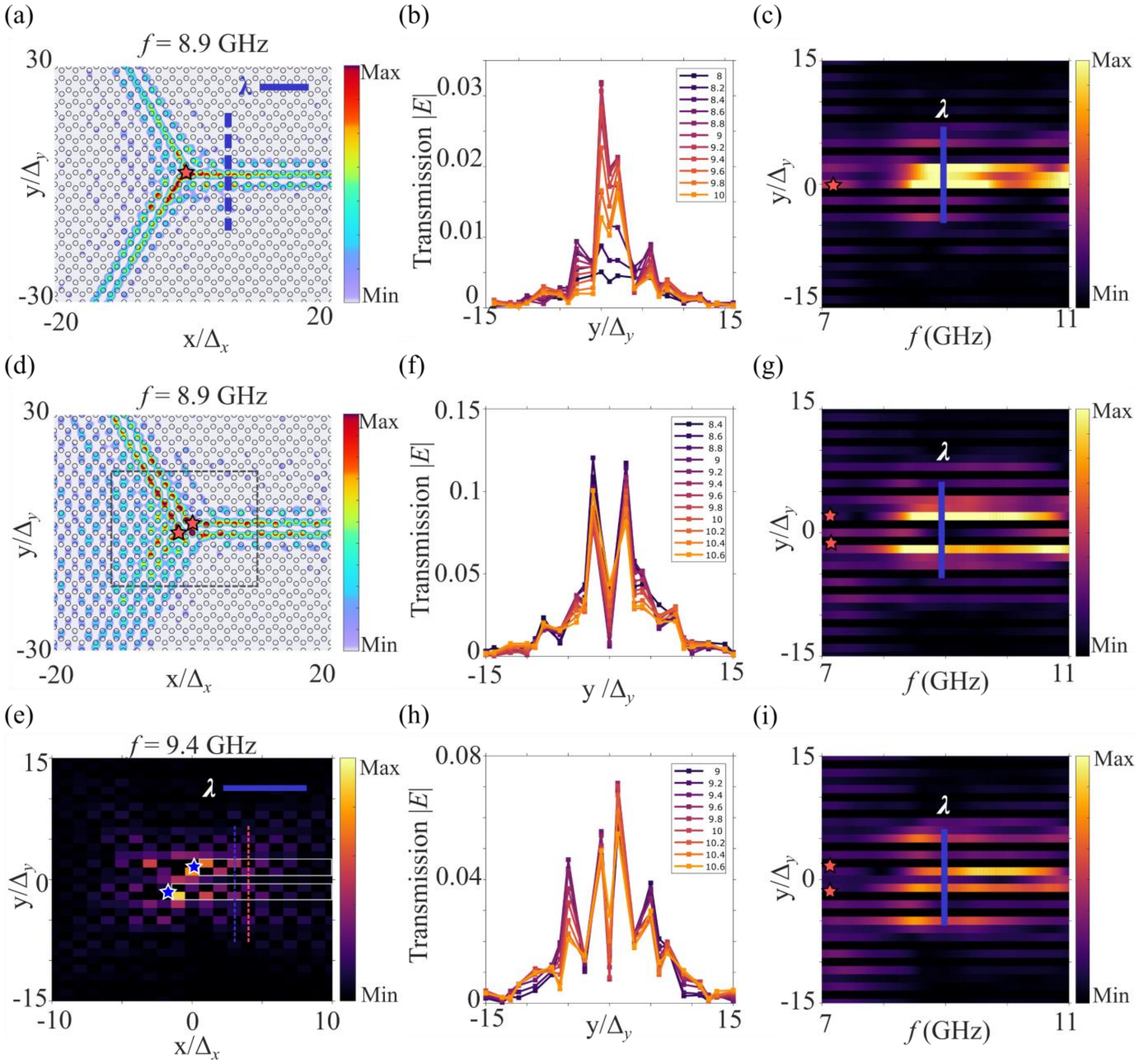

FIG. 3. Experimental demonstration of nonreciprocal subdiffraction imaging. (a) Simulation result on the electric field $|E_z|$ being excited by a point-source at $f$ = 8.9 GHz. (b) The experimentally measured filed distribution along the dashed lines in (a). (c) Spectra view on the experimentally measured beam profile to show subwavelength beam width. A reference scale of wavelength for $f$ = 8.9 GHz is provided. (d) The simulation result on the field distribution for two point-sources at $f$ = 8.9 GHz. (e) The experimentally measured field patten within the dashed box in (d) at $f$ = 9.4 GHz. (f-g) The measured filed distribution along the blue dashed lines in (e), and the corresponding spectra performance of imaging on two point-sources. (h-i) The measured beam profile and spectra view for the pink line in (e).